\documentclass[aps,twocolumn,showlabels,showrefs,amsmath,amssymb,pre,superscriptaddress, floatfix, colors]{revtex4-1}

\usepackage{graphicx}
\usepackage{dcolumn}
\usepackage{bm}
\usepackage{amssymb}
\usepackage{multirow}

\newcommand{\1}{\begin{equation}}
\newcommand{\2}{\end{equation}}
\newcommand{\ea}{\begin{eqnarray}} 
\newcommand{\ee}{\end{eqnarray}}

\begin{document}

\title{Optimal Control Strategies for Active Particle Navigation}
\date{\today}
\author{Benno Liebchen}
\email[]{liebchen@hhu.de}
\thanks{equal contributions}
\affiliation{Institut f\"{u}r Theoretische Physik II: Weiche Materie, Heinrich-Heine-Universit\"{a}t D\"{u}sseldorf, D-40225 D\"{u}sseldorf, Germany}
\author{Hartmut L\"owen}
\email[]{hloewen@hhu.de}
\thanks{equal contributions}
\affiliation{Institut f\"{u}r Theoretische Physik II: Weiche Materie, Heinrich-Heine-Universit\"{a}t D\"{u}sseldorf, D-40225 D\"{u}sseldorf, Germany}
\email[]{liebchen@hhu.de, hloewen@hhu.de}
%

\begin{abstract}
The quest for the optimal navigation strategy in a complex environment is at the heart of microswimmer applications like cargo carriage or 
drug targeting to cancer cells.
Here, we formulate a variational Fermat's principle for microswimmers determining the optimal path 
regarding travelling time, energy dissipation or fuel consumption. For piecewise constant forces (or flow fields), 
the principle leads to Snell's law, showing that the optimal path is piecewise linear, as for   
light rays, but with a generalized refraction law. 
For complex environments, like general 1D-, shear- or vortex-fields, we obtain 
exact analytical expressions for the optimal path, showing, for example, that microswimmers
sometimes have to temporarily navigate away from their target to reach it fastest. 
Our results might be useful to 
benchmark algorithmic schemes for optimal navigation.
\end{abstract}

\maketitle

\paragraph*{\textbf{Introduction}}\label{ra_sec1}
Microswimmers \cite{Ramaswamy2010,Elgeti2015} continuously convert energy into mechanical motion and can self-propel 
in viscous solvents at low Reynolds number. 
Often, they move with an approximately constant speed, but continuously
adapt their swimming direction to accomplish survival tasks.
For algae and spermatozoa \cite{Dusenbery2009}, 
finding an optimal swimming direction decides on their success 
to escape predators and to find prey and mates \cite{Volpe_Volpe_PNAS_2017}. Likewise, the life of bacteria 
rests upon their chemotactic navigation tasks towards food and 
away from toxins \cite{Berg2008,ReviewLL}.
In the flourishing realm of synthetic microswimmers \cite{our_RMP,Menzel2015,Zottl2016,Kurzthaler2018,Lei2018}, in turn, 
controlling the choice of the swimming direction is crucial for technological and 
medical applications like delivering drugs \cite{Popescu_Dietrich_2012,drug_delivery} or other cargo \cite{Baraban2012,Ma2015,Demirors2018,Ghosh2018}
towards a prescribed target. Here, the swimming direction can be controlled via 
external chemical \cite{ReviewLL,Stark2018,Robertson2018,Gonzalez2018}
or electromagnetic fields \cite{Demirors2018} but also by 
feedback-based strategies \cite{Vicsek,Mano2017,Cichos_et_al_Nature_Communications_2018}.

Considering microswimmers with a prescribed deterministic velocity (which may depend on space) and an adjustable self-propulsion direction
in a 2D complex environment, 
here we ask for the optimal path 
to reach a target. 
Contrasting recent (algorithmic) optimization procedures \cite{Haeufle2016,Colabrese2017,Yang2018,Khadem2018,Selmke2018,Nava2018}, 
here we develop a variational approach,  
informing a generalized Fermat's principle for optimal microswimmer navigation, which can be used to calculate the optimal 
path, e.g. regarding 
travelling time, energy dissipation or fuel consumption.

\begin{figure}
  \centering
\includegraphics[width=0.48\textwidth]{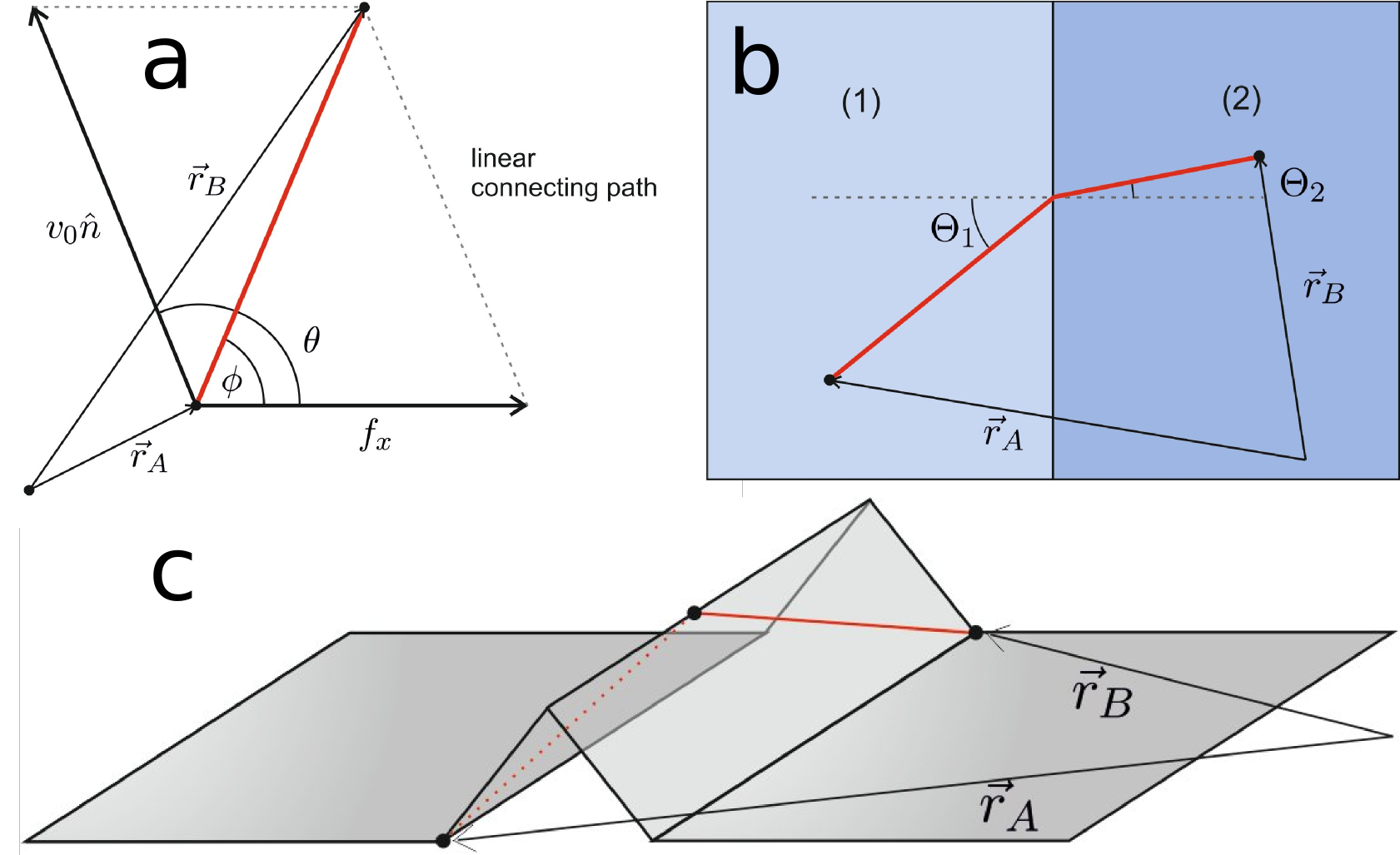}
  \caption{(a) Optimal microswimmer trajectory (red) 
between ${\bf r}_A$ and ${\bf r}_B$ for a constant
flow or force field ${\bf f}=f_x {\bf e}_x$. The optimal orientation ${\bf \hat n}$ 
is fixed by the condition 
that $v_0{\bf \hat n} + f_x {\bf e}_x$ is parallel to ${\bf r}_B - {\bf r}_A$. 
(b,c) Snell's law for microswimmers determining the optimal path in a piecewise homogeneous medium (each with a constant force/flow field and a
specific self-propulsion speed), illustrated for two different 'materials' (b) and a linear ramp potential (c). 
}
\label{Fig:snell}
\end{figure}

Specifically, for vanishing or constant flow and force fields, 
Fermat's principle for microswimmers reduces to its classic counterpart in geometrical optics \cite{Fermat}, showing that microswimmers take the same (straight) path as light rays,
with a speed differing from the bare self-propulsion velocity. 
Consequently, in piecewise linear media, the optimal trajectory follows from a generalized Snell's law, 
assigning refractive angles to a microswimmer's path (Fig.~\ref{Fig:snell}). 

\begin{figure*}
\includegraphics[width=0.9\textwidth]{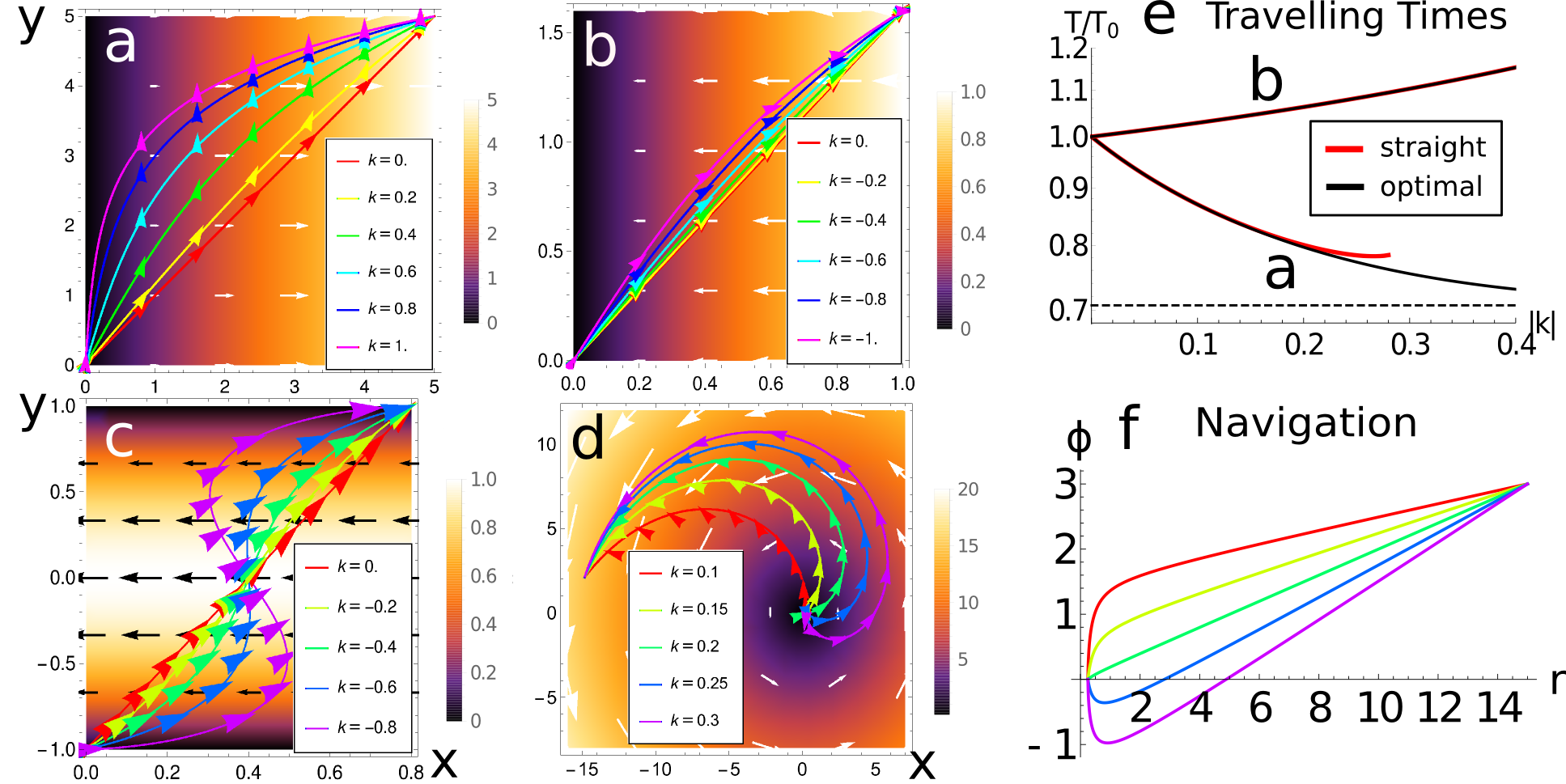}
\caption{Optimal trajectories (lines) and navigation strategy (arrows on lines show ${\bf \hat n}=(\cos\phi,\sin\phi)$) 
in various flow or force fields ${\bf f}$. 
a,b.) 1D linear field ${\bf f}=(k x,0)$;
c.) Shear flow (pipe or plane Poiseuille flow) ${\bf f}=(k[1-y^2/R^2],0)$;
d.) Vortex field ${\bf f}=k(-y,x)$.
Background colors and white arrows show strength and direction of the reduced force ${\bf f}/k$.
e.) Travelling time $T$ (black lines) for problems shown in panels a,b, 
relative to the optimal travelling time $T_0=T(k=0)$. The dashed line represents $T(k\rightarrow 1)/T_0$.
Red lines show $T/T_0$ for a straight trajectory (where existent).
f.) Orientation angle $\phi(r)$ for the trajectories in panel d.
Length and time units are arbitrary, e.g. $\mu m,s$, and $v_0=1$.} 
\label{opttrajs}
\end{figure*}

\begin{figure}
\includegraphics[width=0.44\textwidth]{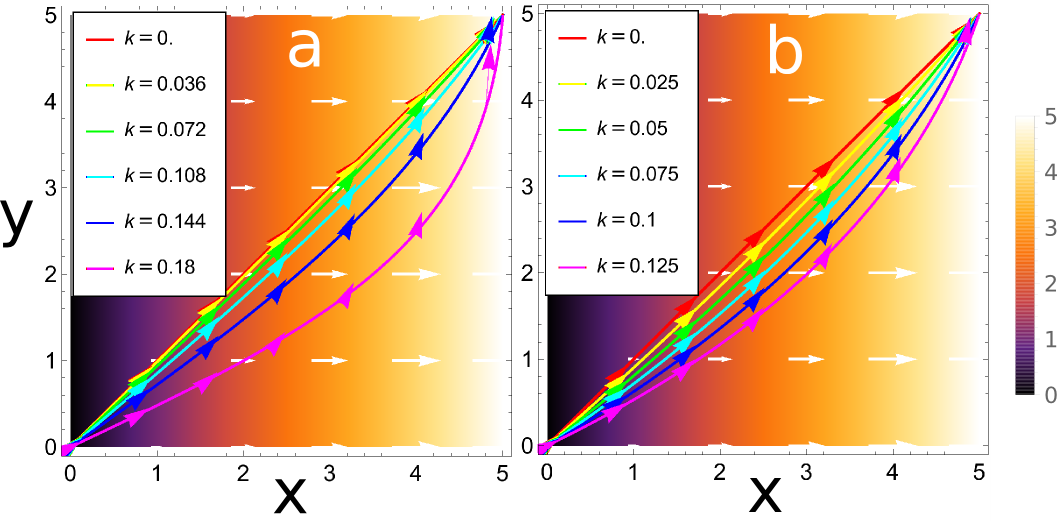} 
\caption{Trajectories minimizing the power dissipated into the fluid (a) and fuel consumption (b). Parameters ${\bf u=0}, {\bf f}=(k x,0)$, 
$v_0=1$ and $\gamma=1$ (a) and $\gamma=1-kx$ in (b). 
Colors, arrows and units as in Fig.~2.}
\label{fig3}
\end{figure}

In complex environments, such as  
general shear-flow problems, isotropic force and vortex-shaped flows and forces, Fermat's principle allows us to calculate 
exact analytical expressions for optimal microswimmer 
trajectories.
These trajectories can have nontrivial shapes (Fig.~\ref{opttrajs}): for instance, a microswimmer 
in a vortex flow field 
sometimes has to swim temporarily away from its target to reach it fastest (Fig.~\ref{opttrajs}). To save fuel, in turn, 
significant excursions as compared to the shortest path can pay off (Fig.~\ref{fig3}).

While some of our results, like the minimization of self-propulsion power, reside in the low Reynolds number world of microswimmers,
those which optimize travelling time, 
might apply even in the macroworld, e.g. to route-planning for airplanes in slowly varying crosswinds  
or to human swimmers aiming to cross a river in minimal time. 
Specifically for such time-optimization problems, our work creates a formal bridge between microswimmer physics and 
Zermelo's classical navigation problem 
\cite{Zermelo1931}, which has been overlooked so far, perhaps because the latter is primarily discussed in the mathematical and engineering literature 
\cite{Zermelo1931,Funk1962,Bryson1975,Cesari1983,Cara1999}.
(Surprisingly, our general solutions for the optimal path might be unknown 
even in that literature \cite{Zermelo1931,Funk1962,Bryson1975,Cesari1983,Cara1999,McShane1937}.) 
\\Our results should be useful for a broad range of microswimmer applications from targeted drug delivery 
\cite{Popescu_Dietrich_2012,drug_delivery} to fuel saving. They might also find applications for 
benchmarking machine learning algorithms applied to optimize navigation \cite{Colabrese2017,Muinos2018} or to 
studies exploring if ocean fish or other swimmers manage to find the path of least resource consumption 
\cite{Hays2014,Mclaren2014}.

\paragraph*{\textbf{Fermat's principle for microswimmers}} 
Consider an overdamped microswimmer (or self-propelled particle) in 2D,
with time-dependent position ${\bf r}(t) =(x(t),y(t))$ 
and orientation ${\bf \hat n}(t) = (\cos \phi (t), \sin \phi (t))$ by:
\begin{equation} \label{eq:II}
 \dot{{\bf r}} =  v_0( {\bf r})  {\bf \hat n} + {\bf f}( {\bf r}); \quad 
\dot {\phi} = M_0(t) 
\end{equation}
Here, $v_0( {\bf r})$ denotes the swimming speed which can be position-dependent
\cite{Lozano_2016,Geiseler_Haenggi_Marchesoni,Maguera_Brendel, Stenhammer} and $ {\bf f}( {\bf r})$ is the overall external field
${\bf f}( {\bf r}) =  {\bf u}( {\bf r}) +  {\bf F}( {\bf r}) / \gamma ( {\bf r})$, with 
${\bf u}( {\bf r})$ and ${\bf F}( {\bf r})$ being external solvent flow and force fields 
and $\gamma ({\bf r})$ being the Stokes drag coefficient, which can also vary spatially (as relevant for viscotaxis 
\cite{PRL_2018_Liebchen_Monderkamp});
$M_0(t)$ is a reduced active torque.
We assume that $M_0(t)$ can be controlled on demand (e.g. v.a external fields) which is equivalent to 
choosing an optimal ${\phi}(t)$.
Here, any external torque or rotational noise in Eq.~(\ref{eq:II}) can be absorbed in $M_0(t)$ and translational noise is neglected
as commonly done for microswimmers.
Given starting and target positions ${\bf r}(t=0) = {\bf r}_A$, 
${\bf r}(t=T) ={\bf r}_B$, we now ask for the optimal
connecting trajectory, which is 
compatible with the equations of motion,
and minimizes the traveling time $T$, for given 
$v_0( {\bf r})$, $ {\bf f}( {\bf r})$. 
This is a
well-posed mathematical variational problem leading to a {\it generalized Fermat's principle for active particles}.

To minimize traveling time, we write 
$T=\int_{x_A}^{x_B} dx \frac{1}{|\dot{x}|}$  
and describe the connecting curve by a function $y(x)$, using $y'(x) = dy/dx$ for its derivative, see Fig.~\ref{Fig:snell}a. 
Then we solve Eq.~(\ref{eq:II}) for ${\bf \hat n}$, square it and express
${\dot{y}} = y'(x) {\dot{x}}$ to arrive at $({\dot{x}} -f_x)^2 + (y'{\dot{x}} -f_y)^2 = v_0^2$. Solving this equation for 
$ {\dot{x}}$ yields 
a functional for $T$
\begin{equation} \label{eq:T_functional} 
T[y(x), y'(x),x] =\int_{x_A}^{x_B} dx \,L(y(x), y'(x),x)
\end{equation}
where we have defined the Lagrangian
\begin{equation} \label{eq:Lagrange_density}
L= \frac{ (1+y'^2)}{\left|f_x + y' f_y \pm \sqrt{v_0^2(1+y'^2) - (f_y-y' f_x)^2} \right|}
\end{equation}
depending on $f_x$, $f_y, v_0$, which are prescribed functions of $x, y$.
Here, the sign leading to the shorter travelling time is the relevant one. 
A necessary condition to minimize $L$ now follows from the 
Euler-Lagrange equation \cite{Goldstein_Classical_Mechanics_textbook}
${{d}\over{dx}} {{\partial L}\over{\partial y'}} - {{\partial L}\over{\partial y}} =0$
yielding a boundary value problem for a second-order differential equation. 
(Specifically for ${\bf f}=(f_x,f_y)={\bf 0}$, we recover Fermat's principle of geometrical optics with 
$v_0( {\bf r})$ replacing the reduced light speed $c_0/n({\bf r})$, 
where $c_0, n({\bf r})$ are the vacuum speed of light and the space-dependent refraction index.)

\paragraph*{\textbf{Snell's law for microswimmers}}
We first consider a microswimmer with constant $v_0$ in a simple environment, given by a gravitational force $m^\ast {\bf g}$ \cite{Enculescu_Stark,Wolff_Stark,ten_Hagen_Bechinger_2014} 
and a constant flow ${\bf u}_0$. 
Choosing an appropriate coordinate system, allows us to write ${\bf f}=f_x {\bf e}_x$
with $f_x =| {\bf u}_0 + m^*{\bf g}/\gamma| = const$, and
the Euler-Lagrange equation reduces 
to a conservation law \cite{Goldstein_Classical_Mechanics_textbook}
\begin{equation} \label{eq:gravity}
{{d}\over{dx}} {{\partial   (1+y'^2)/|  f_x  \pm \sqrt{ f_x^2 + (v_0^2 - f_x^2  )(1+y'^2 )}| }\over{\partial y'}}  =0 
\end{equation}
Thus, $y'(x)$ is constant, i.e. the connecting line between ${\bf r}_A$ and  ${\bf r}_B$ is straight \cite{Zermelo1931}.
To reach its target fastest, the microswimmer thus has to self-propel in a direction ${\bf \hat n}$ such that
${\bf u}_0 + m^*{\bf g}/\gamma + v_0 {\bf \hat n}$ is parallel to ${\bf r}_B - {\bf r}_A$ (Fig.~\ref{Fig:snell}a), yielding
\begin{equation}
\cos\phi=\pm \sqrt{\cos ^2\theta \left[1-\frac{f_x^2}{v_0^2} \sin^2\theta\right]}-\frac{f_x}{v_0} \sin ^2\theta
\end{equation}
where usually the $+$ sign is relevant.
The microswimmer can reach its target if $v_0^2>f_x^2 \sin\theta^2$, where
$\theta$ is the (smallest) angle between ${\bf r}_B - {\bf r}_A$ and ${\bf f}$. 
Its velocity along the trajectory is 
$v_{\rm eff} = |{\bf f}+v_0 {\bf \hat n}|$
and the total traveling time is $T = v_{\rm eff}/|{\bf r}_A - {\bf r}_B|$. 

When ${\bf r}_A,{\bf r}_B$ lie in different homogeneous media, characterized by constant ${\bf f}^{(i)}$ and
$v_0^{(i)}$ ($i=1,2$), and separated by a planar interface 
the optimal trajectory must be piecewise linear (Fig.~\ref{Fig:snell}b). 
(This is because the optimal trajectory between start/target point and intersection point is straight, 
independently of the location of the intersection point.)
The consequence is a generalized Snell's law for microswimmers, with a generalized refraction formula
\begin{equation} \label{eq:Snellius}
{{\sin \Theta^{(1)}}\over {\sin \Theta^{(2)}}} = {{v_{\rm eff}^{(1)}}\over {v_{\rm eff}^{(2)}}}
\end{equation}
where $\Theta^{(i)}$ is the angle between the interface normal and the trajectory in medium $i$.  
The standard Snell-formula emerges for ${\bf f}^{(i)}=0$, whereas $v_{\rm eff}^{(i)}$ generally 
depends on $\Theta^{(i)}$, i.e. (\ref{eq:Snellius}) is an implicit equation.
We illustrate Snell's law and the resulting refraction angles for a microswimmer crossing an 
interface between two fluids in 
Fig.~\ref{Fig:snell}b, and for 
a swimmer 
surmounting a finite and piecewise-linear potential barrier in Fig.~\ref{Fig:snell}c.
Eq.~(\ref{eq:Snellius}) 
applies if ${v_0^{(i)}}^2 > [f_x^{(i)} \sin\theta^{(i)}]^2$ in both media;
if the criterion is violated in one medium, a negative refraction index can arise, as in metamaterials \cite{Shelby2001,Smith2004}.

\paragraph*{\textbf{Complex Environments}}
Let us now explore the optimal path in more generic fields. 
\\\textbf{(i) Exploiting linear flow:}
In the quasi-1D case ${\bf f}=f(x){\bf e}_x$, $ v_0=v_0(x)$, we obtain 
$\partial_{y(x)} L=0$, i.e. $y$ is a cyclic variable, and the Euler-Lagrange equation shows that 
$\partial_{y'(x)}L=c_0$ where $c_0$ is constant along the optimal path. 
Resolving for $y'(x)$ yields (both for $+,-$ in Eq.~\ref{eq:Lagrange_density})
\begin{equation}
y'(x)=\frac{\pm c_0 v_0}{\sqrt{1-c_0^2 (v_0^2 - f^2)}} \label{1dforce}
\end{equation}
which determines the shape of the optimal path for an arbitrary $f(x)$, with $c_0$ and the integration constant
being fixed by the boundary conditions $y(x_A)=y_A$ and $y(x_B)=y_B$.
(Since $\pm$ can be absorbed in $c_0$ both branches of Eq.~(\ref{1dforce}) yield identical boundary value solutions.)
Eq.~(\ref{1dforce}) can be exactly integrated e.g. for 
$f(x)=k x, k/x, k {\rm e}^{\alpha x}$ with $k,\alpha$ being arbitrary (real) constants, 
and otherwise numerically. 
Exemplarily considering $f(x)=k x$ 
(Fig.~\ref{opttrajs}a), we recover the straight line for $k=0$; as  
$k$ increases, the optimal trajectory
increasingly bends away from the straight line.  
To understand how such a detour pays off regarding travelling time, 
consider the 
$k=1$-case: here, the microswimmer self-propels in $y$-direction only, whereas the external field 
generates all required motion in 
$x$-direction. That way, the 
travelling 
time reduces by a factor of $\sqrt{2}$ as compared to the straight trajectory at $k=0$. 
If $k < \sqrt{2}v_0/5$, the microswimmer can alternatively reach its target by following the geometrically shortest, straight path,
i.e. to minimize travelling distance rather than time. 
Comparing travelling times
(Fig.~\ref{opttrajs}e) shows that the straight-line motion is never optimal for $k\neq 0$, but 
only marginally worse than the optimal one for most relevant $k$-values.  
Thus, for microswimmers seeing only their local environment, a very simple, yet sensible strategy could be to always head straight towards the target.
This strategy works even better in our next example. %
\\\textbf{(ii) Optimal navigation in upwards flow direction:} 
A swimmer 
aiming to reach a target located in upwards
flow (force) direction (Fig.~\ref{opttrajs}b), benefits from  
staying 'above' the straight line. This helps the swimmer to avoid strong opposing flow regimes unnecessarily early, 
but makes the resulting path longer. 
The optimal compromise is a path slightly above the diagonal, following which requires the 
swimmer to steer increasingly against the flow. 
(This agrees with Zermelo's qualitative finding \cite{Zermelo1931,Cara1999} that the steering
``must always be toward the side which makes the wind component acting against the steering direction larger'').
The optimal path again reduces travelling time as compared to the straight line (Fig.~\ref{opttrajs}e), but only very slightly, 
showing once more, that moving straight towards a target serves as an excellent alternative strategy. 
\\\textbf{(iii) Crossing a pipe:}
Analogously to our previous calculation, we obtain an exact expression for the optimal path for a general 
shear-flow problem \cite{ten_Hagen_2011,Tarama_2013} ${\bf f}=f(x){\bf e}_y$ ($v_0=v_0(x)$), where $+,-$ in Eq.~(\ref{eq:Lagrange_density})
both yield (modulo an irrelevant sign of $c_0$): 
\begin{equation}
y'(x) = \pm \frac{c_0 v_0^2 + f - c_0 f^2}{v_0\sqrt{(c_0 f -1)^2-c_0^2 v_0^2}}
\end{equation}
Here the $+$ branch is the relevant one in all examples we have explored. 
Let us illustrate this result for a microswimmer aiming to cross a pipe
${\bf f}=k[1-x^2/R^2]{\bf e}_y$ (planar Poiseuille flow); see Fig.~\ref{opttrajs}c. Here, 
to reach its target fastest, the microswimmer takes an increasingly 
S-shaped path, as $-k$ increases. 
In particular, 
to cross the pipe most efficiently
in upwards flow direction, the microswimmer is obliged to temporarily move down the flow. 
(For $k\lesssim -0.82$ the target is unreachable.)
\\\textbf{(iv) 2D environments:}
To explore the optimal path in 2D force and flow fields, 
as created e.g. by a rotating bucket or
an optical trap \cite{Szamel_PRE,Pototsky,volpe2013,nourhani2015,ribeiro2016}, 
we rederive the Lagrangian $L=L(r,\phi(r),\phi'(r))$ in polar coordinates $(r,\phi)$ parameterized by $r$, 
for ${\bf f}(r,\phi)=f_r(r,\phi) {\bf e}_r + f_\phi(r,\phi) {\bf e}_\phi$ where 
${\bf e}_r = (\cos\phi,\sin\phi)$ and 
${\bf e}_\phi = (-\sin\phi,\cos\phi)$: 
\begin{equation}
L=\frac{1+r^2 \phi'^2(r)}{\left| f_r + r \phi' f_\phi \pm \sqrt{v_0^2 - f_\phi^2 + r \phi'[2 f_r f_\phi + r \phi' (v_0^2 - f_r^2)]} \right|} \label{Lpolar}
\end{equation}
For isotropic forces $f_r=f(r); f_\phi=0$ (like the simplest optical traps) and $v_0=v_0(r)$,
we exploit that 
$\partial_{\phi(r)} L=0$, so that the Euler-Lagrange equations yield $\partial_{\phi'(r)}L=c_0$ with $c_0$ being constant again.
Hence, the optimal trajectory for an arbitrary isotropic potential reads (both for $+,-$ in Eq.~\ref{Lpolar})
\begin{equation}
\phi'(r) = \frac{c_0 v_0}{\sqrt{r^4+c^2_0 r^2 [f^2-v_0^2]}}
\end{equation}
Similarly, for vortex fields $f_r=0; f_\phi=f(r); v_0=v_0(r)$ we find ($+,-$ sign in Eq.~\ref{Lpolar} again lead to the same two solutions, modulo the 
irrelevant sign of $c_0$)
\begin{equation}
\phi'(r) = \pm\frac{c_0 v_0^2 + r f - c_0 f^2}{r v_0 \sqrt{r^2 - c_0^2 v_0^2 - 2 c_0 r f + c_0^2 f^2}}
\end{equation}
To exemplify these results, consider 
a microswimmer in the center of a rotating flow ${\bf f}=k(-y,x)=k r {\bf e}_\phi$ in a (nonrotating) bucket aiming to reach a specific point on the bucket rim as soon as possible.
As shown in Fig.~\ref{opttrajs}d, reaching the target fastest, sometimes obliges the swimmer to initially moves away from it (cases $k=0.2; 0.25; 0.3$). 
Here, the swimmer's orientation strongly changes at small $r$ only (panel f), where $f$ is weak; i.e. the swimmer performs its navigation task at small $r$, letting 
the flow advect it to the target afterwards.  

\paragraph*{\textbf{Optimizing drag power}} 
To illustrate path optimization regarding quantities different from $T$, 
we first define the drag power dissipated into the fluid as $P=\gamma (\dot {\bf r}-{\bf u})^2$, simplifying to 
$P=\gamma |\dot x|^2 [1+y'(x)^2]$ for ${\bf u=0}$.
Analogously to our previous approach, 
we write the energy $E$ dissipated along a microswimmers' path $y(x)$ into the solvent as (still for ${\bf u=0}$)
\begin{equation}
E=\int {\rm d}t P(t) = \int\limits_{x_1}^{x_2} dx L_{\rm P}; \quad L_{\rm P} = \frac{\gamma (1+y'^2)}{L(x,y,y')}
\end{equation}
where $\gamma,v_0,{\bf F}$ may depend on ${\bf r}$.
Following the Euler-Lagrange equation for $L_{\rm P}$ shows that 
$L_{\rm P}$ has the same cyclic variables as $L$, allowing us to follow our earlier solution strategy. 
Specifically for 
1D fields ${\bf F}/\gamma=f(x){\bf e}_x$, the path minimizing $E$ 
is determined by (both for $+,-$ in Eq.~\ref{eq:Lagrange_density})
\begin{equation}
y'(x)=\frac{c_0 v_0}{\sqrt{(f^2-v_0^2)[c_0^2+\gamma^2 (f^2-v_0^2)]}}
\end{equation}
where $v_0,\gamma,f$ may all depend on $x$ and where $c_0$ and the integration constant are again fixed by boundary conditions.
Exemplaric trajectories for $f=k x$ (Fig.~\ref{fig3}a) show that minimizing energy dissipation requires a microswimmer to 
take a path of opposite curvature as compared to the fastest one (Fig.~\ref{opttrajs}a). 
Physically, the microswimmer compromises between minimizing travelling distance and 
avoiding regions of strong force, since moving in force direction is costly, since 
$P \propto ({\bf \hat n} v_0 + {\bf f})^2$.
(Notice, that for ${\bf u}\neq {\bf 0}, {\bf F=0}$
the drag power simplifies to the self-propulsion power $P=\gamma v_0^2$, discussed next.) 

\paragraph*{\textbf{Fuel Saving}} 
Finally, we minimize the self-propulsion power $P=\gamma v_0^2$ integrated along the path, assumed to be  
proportional to the fuel required. 
Here, if either ${\bf u}={\bf 0}$ or ${\bf F=0}$ the relevant Lagrangian reads $L_{\rm SP}=\gamma v_0^2 L$.
For instance, when ${\bf F}=f(x){\bf e}_x$ and $\gamma,v_0$ depends on $x$ only, the path minimizing fuel consumption 
is determined by
\begin{equation}
y'(x)=\frac{c_0 v_0}{\sqrt{c_0^2 (f^2-v_0^2)+v_0^4 \gamma^2}} 
\end{equation}
The resulting path is identical to the one minimizing $T$ if
$v_0^2 \gamma$ is constant ($v_0^4 \gamma^2$ can be absorbed in $c_0$), but not in general. In fact, optimizing fuel consumption 
sometimes requires microswimmers to make significant excursions; e.g. for ${\bf f}=k x {\bf e}_x$ and $\gamma=1-k x$ microswimmers initially 
navigate towards low viscosity regions before increasinly turning towards the target (Fig.~\ref{fig3}b) 

\paragraph*{\textbf{Conclusions}}
Fermat's principle for microswimmer navigation
connects active matter with geometrical optics and optimal control theory to determine 
the optimal strategy to reach a target e.g. in minimal time or with minimal fuel consumption. 
Our exact and general results for microswimmers in 
1D, shear and vortex fields can be used to benchmark approximative schemes for optimal navigation, including 
machine-learning-based ones \cite{Colabrese2017,Muinos2018} and perhaps also to test 
the extend evolution has optimized swimming paths of sea animals \cite{Hays2014,Mclaren2014}.

Future work could generalize our approach to 3D \cite{Wysocki_Gompper}, 
viscoleastic solvents \cite{viscoelastic}, associated intertial effects \cite{Dauchot_Frey_PRL_2013} or 
curved manifolds \cite{Skepnek_Henkes1,Skepnek_Henkes2,Janssen_Kaiser_Loewen}, possibly linking microswimmer physics 
with geodesics in the curved space-time of general relativity, and should of course account for   
Brownian noise \cite{Baer,Baskaran,Haenggi},
where the Onsager-Machlup formulation \cite{Onsager_Machlup,Asheichyk2018} might provide a formal
link to quantum mechanics.

\paragraph*{Acknowledgments}
We thank C. Scholz and A. Ivlev for helpful discussions and F. Hauke for preparing Fig. 2. HL
gratefully acknowledges support by the Deutsche
Forschungsgemeinschaft (DFG) through
LO 418/19-1.

\end{document}